\documentclass[aps,prl,showpacs,amsmath,amsfonts,superscriptaddress,twocolumn]{revtex4}
\usepackage{graphicx}% Include figure files
\usepackage{dcolumn}% Align table columns on decimal point
\usepackage{bm}% bold math
\usepackage{pslatex}
\begin{document}

\title{Ultracold atoms in one-dimensional optical lattices \\ 
       approaching the Tonks-Girardeau regime}

\author{L. Pollet}
\email{Lode.Pollet@UGent.be}
\author{S. M. A. Rombouts}
\affiliation{Vakgroep Subatomaire en Stralingsfysica, Universiteit Gent,
  Proeftuinstraat 86, 9000 Gent, Belgium}
\author{P. J. H. Denteneer}
\affiliation{Instituut-Lorentz, Universiteit Leiden, P.O. Box 9506, 
             2300 RA Leiden, The Netherlands}

\date{\today}
\begin{abstract}
Recent experiments on ultracold atomic alkali gases in a 
one-dimensional optical lattice have demonstrated the transition 
from a gas of soft-core bosons to a Tonks-Girardeau gas 
in the hard-core limit, where one-dimensional bosons 
behave like fermions in many respects.
We have studied the underlying many-body physics
through numerical simulations which accommodate both
the soft-core and hard-core limits in one single framework.
We find that the Tonks-Girardeau gas is reached only 
at the strongest optical lattice potentials.
Results for slightly higher densities,
where the gas develops a Mott-like phase already
at weaker optical lattice potentials,
show that these Mott-like short range correlations
do not enhance the convergence to the hard-core limit.
\end{abstract}


\maketitle
{\it Introduction.} --- Since the prediction by Jaksch 
{\it et al.}~\cite{Jaksch98} on the experiment by Greiner 
{\it et al.}~\cite{Greiner02}, ultracold atoms in optical lattices have 
been the focus of much activity. 
By tightly confining the motion in the transverse direction,
an array of quasi one-dimensional optical lattices results~\cite{Stoferle04},
where particle exchange between the one-dimensional tubes is suppressed.
The role of quantum fluctuations is enhanced in one dimension compared 
to the three-dimensional case, such that traditional mean-field theories fail. 
Instead, the long-range low-energy physics is described 
by the Luttinger liquid model. 
In the limit of infinite repulsion between the atoms the atomic gas is called a
Tonks-Girardeau gas~\cite{Girardeau60,Lieb63} (TG). 
Because of the blocking of double occupancies, the resulting hard-core bosons 
have some properties very similar to non-interacting fermions:
e.g. the density profiles become indistinguishable.
However, more complicated properties such as the momentum distribution
remain discriminating characteristics~\cite{Paredes04}.
The regime of strong repulsions between bosons
has been studied experimentally~\cite{Laburthe04} and
theoretically~\cite{Astrakharchik03} for an atomic gas not subject to an
optical potential, but the acquired values for the ratio 
of the repulsive interaction strength to the kinetic energy were rather low 
and the TG regime was not seen. 
By using an optical lattice, much higher values for this 
ratio could be reached~\cite{Paredes04}. 
The interpretation of these experiments is complicated by the
finite-size effects due to the harmonic trap.
But even in the homogeneous case, an accurate theoretical
description of the transition from a weakly interacting Bose gas 
to a strongly interacting Tonks gas has to rely on numerical simulations.
Our aim is to model the experimental results of Ref.~\cite{Paredes04}
using one single numerical framework which accommodates both the
weakly and the strongly interacting regime.

The physics of ultracold atoms in optical lattices can be described 
by the Bose-Hubbard model~\cite{Jaksch98}, 
which considers bosons occupying Wannier orbitals. 
The validity of this model is confirmed by the ratio of the central
to first Bragg peak in the experimentally observed momentum distributions, 
which depends only on the shape of the Wannier orbitals (see below).
The TG regime is characterized by the absence of double occupancies
in the many-boson wave function. To identify the TG regime unambiguously,
one has to evaluate whether the experimental results are better described
by soft-core bosons with a considerable overlap or by hard-core bosons
for which double occupations are explicitly suppressed.
Exact results for realistic parameters over the entire range of the axial
optical lattice depths used in the experiment are obtained using quantum Monte
Carlo methods.
We find that the results for soft- and hard-core bosons do not coincide 
except for the strongest optical potentials used in the experiments,
in contrast with the fermionization approach of Ref.~\cite{Paredes04}
which assumes hard-core bosons at all optical-potential strengths.

{\it One-dimensional optical lattice.} --- 
When an ultracold Rb gas of atoms is cooled and loaded into an optical
lattice~\cite{Greiner02} with very tight transverse confinement, its dynamics
is governed by the one-dimensional Hamiltonian, 
\begin{equation}
  H = \frac{p^2}{2m} + V_0(x) + V_T(x) 
     + g_{int} \sum_{i<j} \delta(x_i - x_j),\label{eq:hamiltonian}
\end{equation}
with $m$ the atomic mass, $x_i$ the position of atom $i$, 
$V_0(x) = V_0\sin^2(kx)$ the optical potential 
($V_0$ takes the laser intensity and the dynamic polarizability 
 of the atoms into account)
and $V_T(x)$ the harmonic trapping potential, 
which varies slowly compared to the optical potential. 
The wave vector $k$ of the laser along the axial direction
defines the length scale $\lambda/2$ through $k = 2\pi/\lambda$ 
and the recoil energy $E_R = \hbar^2k^2/(2m)$ which we will use as an
energy scale. 
The Hamiltonian Eq.(\ref{eq:hamiltonian}) reduces to the exactly solvable
Lieb-Liniger~\cite{Lieb63} model for $V_T(x) = 0$ and $V_0(x) = 0$  while it
reduces to a Mathieu equation for $V_T(x) = 0$ and $g_{int}=0$. The
interaction $g_{int}$ between the atoms is determined by the three dimensional
scattering length $a_s$ of the atoms.  Olshanii~\cite{Olshanii98} studied the
scattering problem of two particles in tight waveguides and found for the
effective one-dimensional coupling constant 
\begin{equation}
g_{int} = \frac{2 \hbar^2 a_s}{ma_{\perp}^2} \frac{1}{1-1.033 a_s/a_{\perp}},
\end{equation}
where $a_{\perp} = \sqrt{\hbar/m\omega_{\perp}}$ is the characteristic length
of the transverse harmonic confinement.
For very tight radial confinement it suffices 
to integrate over the $y$ and $z$ directions assuming harmonic confinement, 
yielding $g_{int} = \frac{2 \hbar^2 a_s}{ma_{\perp}^2}$.  
The Wannier orbitals are calculated for the periodic potential 
given by the kinetic and the optical terms in Eq.(\ref{eq:hamiltonian}),
restricted to the lowest band~\cite{Vanoosten01}.
For low-density gases we can express the Hamiltonian of
Eq.(\ref{eq:hamiltonian}) in the Wannier basis, resulting in a  
Bose-Hubbard model~\cite{Fisher89,Jaksch98},
\begin{equation}
H = -J \sum_{\langle i,j \rangle} b^{\dagger}_i b_j + \sum_i \epsilon_i n_i +
\frac{U}{2} \sum_i n_i(n_i - 1),
\label{eq:hubbard_hamiltonian}
\end{equation}
where
the first summation runs over nearest neighbors only, 
the operator $n_i$ counts the number of bosons at site $i$
and the effective parameters $U, \epsilon_i$ and $J$ 
represent the strengths of the on-site repulsion, 
the harmonic trapping and the kinetic hopping, respectively. 
Recent studies of the one-dimensional Bose-Hubbard model mainly focused 
on the Mott-superfluid transition, using a wide range of methods:  
a slave-boson approach~\cite{Dickerscheid03}, 
the numerical renormalization group~\cite{Pollet03}, 
the density matrix renormalization group~\cite{Kollath04}, 
the time-evolving block decimation method~\cite{Clark04} 
and Monte Carlo methods~\cite{Wessel04}.  
The main uncertainties in the model relate to the accuracy 
of the scattering length $a_s$ 
and the renormalization of the effective parameters of the Bose-Hubbard model. 
As we work in the grand-canonical ensemble, the chemical potential $\mu$ 
must be fine-tuned such that the expected number of particles corresponds 
to the experimental number of particles. 
The Bose-Hubbard model is simulated using the stochastic series 
expansion method~\cite{Sandvik99} (SSE)
with locally optimized directed loop updates~\cite{Syljuasen02,Pollet04}. 
From this Monte Carlo simulation thermodynamic observables
such as the energy, the (local) density, the (local) compressibility 
and the one-body correlation function
can be computed exactly in a statistical sense~\cite{Dorneich01} .

\begin{figure}
\includegraphics[width=8.5cm]{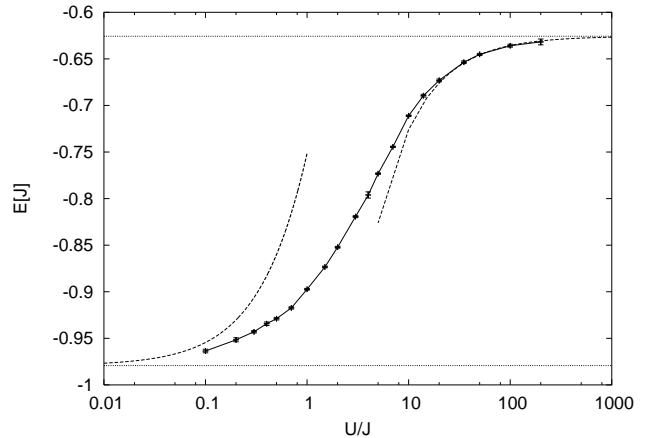}
\caption{
 Internal energy per site as a function of $U/J$ for a homogeneous model 
 with $128$ sites at a temperature $T/J=0.2$. The data points with error bars
 connected by the full line are the energies obtained by the SSE method.
 The energies for non-interacting fermions (upper dotted horizontal line) 
 and non-interacting bosons (lower dotted horizonal line) are shown, 
 together with the Bogoliubov approximation for bosons (dashed line on the
 left) and a first order perturbation theory for fermions (dashed line on the
 right). 
\label{fig:scurve} }
\end{figure}

{\it Homogeneous system.} --- First, we consider a homogeneous 
($\epsilon_i = 0$) atomic gas in a lattice of $L=128$ sites 
with periodic boundary conditions at a very low but finite 
temperature, $T/J = 0.2$. 
In Fig.~\ref{fig:scurve} we show the internal energy per site of this system
for increasing values of $U/J$, keeping the average density fixed at $\langle
n \rangle \approx 0.5$. An ideal Bose gas occurs in the limit of vanishing $U$, 
which is indicated by the lower horizontal line in Fig.~\ref{fig:scurve},
while the ideal Fermi gas is found for $U \to \infty$ 
and indicated by the upper horizontal line. 
For very large values of $U/J$, no site of the lattice will be
doubly occupied and one can apply the Jordan-Wigner transformation 
to map the bosons onto fermions~\cite{Sachdev99}. 
For small but finite $U$ ($U/J \ll 0.1$ in Fig.~\ref{fig:scurve}) , the system
is adequately described by the standard Bogoliubov
approximation~\cite{Vanoosten01}.  
For large $U$ a perturbation of interacting fermions 
was derived in Ref.~\cite{Cazalilla03PRA} to order $1/U$. 
From the log scale in Fig.~\ref{fig:scurve} it appears that the limit 
of non-interacting fermions is reached slowly for values of $U/J>10$. 
For $U/J \sim 1$ one has to resort to numerical methods, 
and we see that the SSE method remains efficient over the entire $U/J$ range. 
Higher temperatures lead qualitatively to the same results, 
but the description in terms of fermions is only valid for higher values of
$U/J$. 
Temperature can be seen as a source for exciting double occupancy on a
particular site, whose likeliness must be suppressed by a stronger on-site
repulsion term.

\begin{figure}
\includegraphics[width=8.5cm]{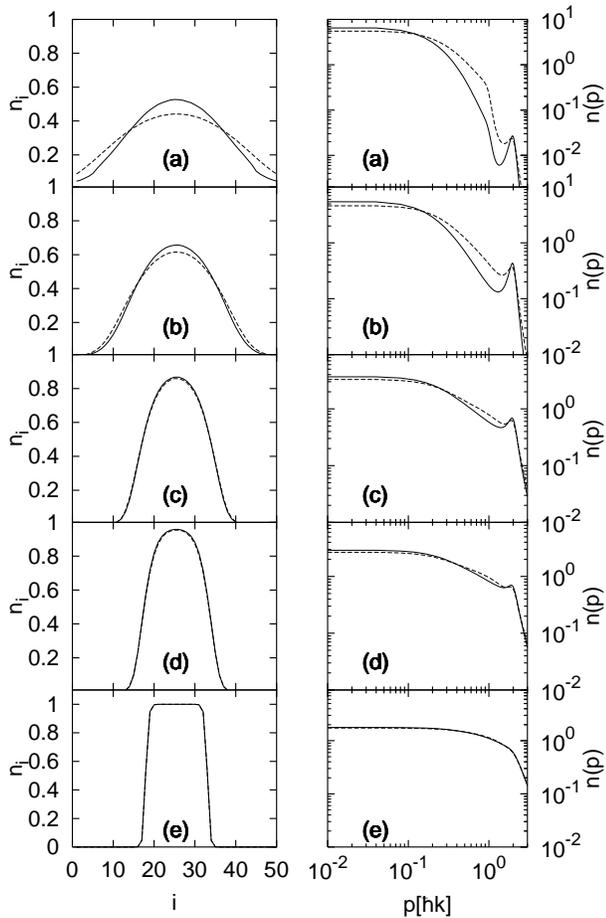}
\caption{
Local densities $n_i$ in coordinate space 
as a function of the site index $i$ (left)
and the corresponding momentum profiles $n(p)$ 
as a function of the momentum $p$ (in units $\hbar k$) on the right. 
The axial optical lattice depths, 
the ratios $U/J$ and the values of the slope parameter $\alpha$ for soft-core
(full line) and $\alpha^{\prime}$ for hard-core bosons (dashed line) are 
(a) $V_0/E_R =    1, U/J =   1.75, \alpha = 2.71, \alpha^{\prime} = 1.69$, 
(b) $V_0/E_R =    5, U/J =   7.85, \alpha = 1.92, \alpha^{\prime} = 1.38$, 
(c) $V_0/E_R =  9.5, U/J =   28.6, \alpha = 1.00, \alpha^{\prime} = 0.78$, 
(d) $V_0/E_R =   12, U/J =  52.28, \alpha = 0.72, \alpha^{\prime} = 0.56$, 
(e) $V_0/E_R =   20, U/J = 258.54, \alpha = 0.33, \alpha^{\prime} = 0.32$. 
In each plot the average number of particles is $\langle N \rangle \approx
15$, the temperature is $T/J= 1$, and the lattice consists of 50 sites.
\label{fig:comp} } 
\end{figure}

{\it Inhomogeneous system.} --- 
The harmonic trapping potential breaks the homogeneity of the system.
For the parameters we follow Ref.~\cite{Paredes04}:
the scattering length $a_s$ of Rb atoms is taken 
to be $a_s = 102(6) a_0$~\cite{Buggle04}, with $a_0$ the Bohr length;
the characteristic length $a_{\perp}$ of the tight confinement 
in the $y$ and $z$ direction is $a_{\perp}= 57.6$nm;
the parameter $\epsilon_i$ in Eq.(\ref{eq:hubbard_hamiltonian}), 
characterizing the trapping in the axial direction, is given by
$\epsilon_i = \int dx V_T(x) |\Psi(x-x_i)|^2 
            \simeq 8 \times 10^{-4}E_R(i - \frac{L}{2})^2$,
with $\Psi(x-x_i)$ the Wannier function centered around site $i$
and $L = 50$ the total number of sites. 
The ratio $U/J$ can be varied by changing the optical potential strength $V_0$.
The temperature $T$ and the number of particles in one tube
are not directly accessible experimentally. 
The averaging over an array of one-dimensional tubes in Ref.~\cite{Paredes04} 
can be understood as an averaging over condensates with
different temperatures and particle numbers. 
However, one can understand the onset of the TG limit from simulations 
for a single tube with a fixed temperature.
We used $T/J=1$ at all interaction strengths, which is 
of the same order as the temperatures estimated in Ref.~\cite{Paredes04}.
In Fig.~\ref{fig:comp} the local densities and the momentum profiles 
are shown for several values of the optical potential strength $V_0$,
in line with the actual values used in the experiment of Ref.~\cite{Paredes04}.
All Monte Carlo simulations consist of at least $20$ chains 
of $2^{16}$ samples with each $50-200$ off-diagonal updates 
such that error bars are not visible. 

Momentum profiles are experimentally measurable 
and can be calculated from a numerical simulation as
\begin{equation}
n(p) = |\Phi(p)|^2 \sum_{j,l} e^{-ip(j-l)} \langle b^{\dagger}_j b_l \rangle,
\end{equation}
where the envelope $\Phi(p)$ is the Fourier transform 
of the Wannier function $\Psi(x)$, 
$p$ denotes momentum in units of $\hbar k$ 
and $\langle b^{\dagger}_j b_l \rangle$ is the
one-body density matrix of the Bose-Hubbard model. 
In Fig.~\ref{fig:comp}, the peak observed at $p = 2 \hbar k$ is the
first-order diffraction peak reflecting the presence of the optical lattice.
The ratio between the height of the central peak and the first-order peak 
is solely related to the width of the Wannier orbitals 
and is not affected by averaging over the array of tubes 
or by the dynamics of the Bose-Hubbard model.
The procedure to calculate the Wannier orbitals outlined above 
yields ratios in good agreement with the experimental data shown 
in Fig.2 of Ref.~\cite{Paredes04}. This suggests that the ramping down along
the axial direction in the experiment proceeded adiabatically, and it
demonstrates that the discrete Bose-Hubbard model is a valid approach to
describe the physics of ultracold atomic alkali gases in optical lattices. 

In every subplot (a-e) of Fig.~\ref{fig:comp} there is a region where
the slope of the momentum distribution is almost linear (on log-log scale),
similar to what occurs in an infinite homogeneous Tonks gas at $T=0$,
which has an infrared divergence $n(p)\propto p^{-1/2}$ at low momenta and 
an asymptotic tail $n(p) \propto p^{-4}$ 
at high momenta~\cite{Olshanii03}.
In our case, the periodicity of the optical lattice sets an upper
momentum scale $p_L=\hbar k$.
The width of the Wannier orbitals sets another upper scale,
$p_W \simeq (V_0/E_R)^{1/4} p_L$, which turns out 
to be larger than $p_L$ for the parameter regimes considered here.
The harmonic trap sets a lower momentum scale 
$p_T=(m \hbar \omega)^{1/2} \simeq 0.1 \hbar k$,
below which the momentum distribution is flattened because 
of the suppression of long-range correlations. 
Due to the trapping potential, the influence of temperature
on the momentum distribution will be different from the
homogeneous case: thermal fluctuations will occur
at the edges of the cloud and therefore they will mainly 
affect the momentum distribution below $p_T$.
Only the momentum distribution in the region 
between $p_T$ and $p_L$ relates directly to the 
short-range dynamics of the Bose-Hubbard model
and might show a power-law behavior similar to the homogeneous system.
We have fitted the linear parts of the log-log curves in this region 
with a power law $n(p) \propto p^{-\alpha}$.
The slope $\alpha$ is sensitive to temperature, density and interaction 
strength, but to first order independent of the Wannier orbitals. 

\begin{figure}
\includegraphics[width=8cm]{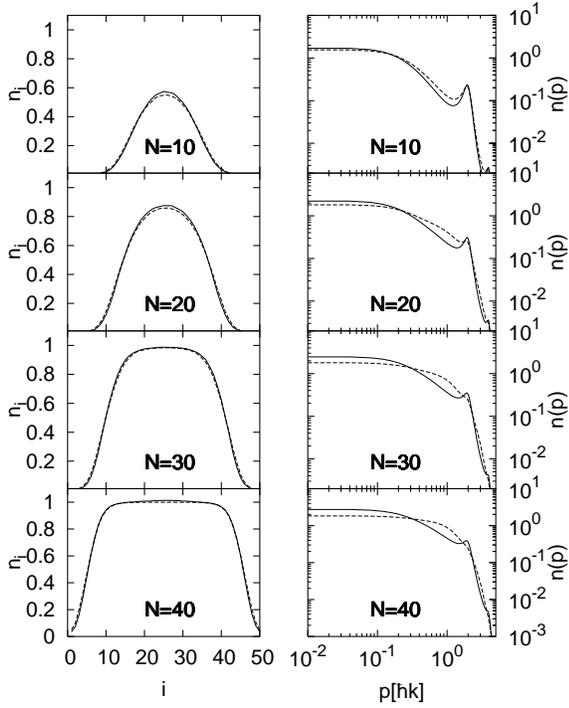}
\caption{Density profiles $n_i$ and momentum density profiles $n(p)$ 
 ($p$ in units $\hbar k$)
  for different fillings $N$ at an optical lattice potential $V_0/E_R = 7$
  and temperature $T/J=1$, for soft-core (full line) 
  and hard-core bosons (dashed line). 
   \label{fig:densities} }
\end{figure}

By comparing the results for soft-core and hard-core bosons
in Fig.~\ref{fig:comp}, one sees that the TG regime is approached 
for optical potentials $V_0/E_R = 9.5$ and $V_0/E_R = 12$, 
while it is fully reached only at $V_0/E_R = 20$,
which in our model corresponds to a ratio $U/J = 259$.
These values are in good agreement with Fig.~\ref{fig:scurve},
where the energy for $U/J = 200$ is only $4\%$ lower 
than the energy of an ideal Fermi gas. 
For lower optical potentials, finite boson-boson interactions 
certainly  need to be taken into account 
and double occupancies in the center of the trap do play an important role.
We see in subplot (e) of Fig.~\ref{fig:comp} that a Mott-like region is formed 
in the center of the trap.
For a homogeneous system in the Mott phase,
the dispersion relation of the excitations has a gap of order
$U$~\cite{Fisher89}, meaning that the role of double occupancies is strongly
suppressed~\cite{Pollet03}. The Mott phase is entered at a ratio $U/J$ as low 
as $1.67$ at $T=0$ for a density of one particle per site~\cite{Kuhner99}.
For the inhomogeneous system, 
the insulating behavior translates into a local compressibility
that tends to vanish in the center of the trap~\cite{Batrouni02}.
Hence, for $T \ll U$ hard-core bosonic behavior can be reached 
for local densities varying from $\langle n_i \rangle = 0$ 
to $\langle n_i \rangle = 1$.
However, the reduced local compressibility does not mean 
that for higher densities the TG regime would be reached 
at weaker optical potential strengths.
Fig.~\ref{fig:densities} shows that a significant difference
between the soft-core and hard-core momentum profiles persists
even if the density-profile develops a Mott-like region,
at an intermediate optical lattice strength $V_0/E_R = 7$ $(U/J=14.3)$.
This indicates that the short-range correlations in the Mott-like region
differ significantly from the short-range correlations in the TG regime.

In conclusion, we have shown that the experiment of Ref.~\cite{Paredes04} is 
very well described by a Bose-Hubbard model based on Wannier orbitals.
Soft-core boson wave functions with a significant contribution of double
occupancies can explain the experimental results over the largest part 
of the optical potential parameter range.  
Only for very deep optical lattices ($V_0/E_R = 20$) 
do the atoms behave as hard-core bosons 
and does the Tonks-Girardeau picture apply. 
The averaging over the array of one-dimensional tubes has only a minor effect 
and does not significantly alter the momentum profiles. 
At higher densities, Mott-like correlations might develop,
but they do not enhance the convergence to the Tonks-Girardeau regime.

We wish to acknowledge fruitful discussions with H.T.C. Stoof, K. Heyde 
and B. Paredes. 
This research was supported by the Research Board of the University of Ghent
and the Fund for Scientific Research - Flanders (Belgium).

\end{document}